\documentclass[10pt,conference,twocolumn]{IEEEtran}
\usepackage{filecontents}
\usepackage[noadjust]{cite}
\usepackage{caption}
\usepackage{amsfonts}
\usepackage{subcaption}
\usepackage[dvips]{color, graphicx}
\usepackage[cmex10]{amsmath}
\usepackage{pgfplotstable}
\usepackage{array}
\usepackage{booktabs}
\usepackage{times}
\usepackage[normalem]{ulem} 

\DeclareMathOperator*{\self}{SELF}
\DeclareMathOperator*{\symm}{SYMM}
\DeclareMathOperator*{\nc}{NC}
\DeclareMathOperator*{\nf}{NF}
\DeclareMathOperator*{\nm}{NM}
\DeclareMathOperator*{\partition}{Par}
\DeclareMathOperator*{\row}{R}
\DeclareMathOperator*{\T}{T}

\long\def\symbolfootnote[#1]#2{\begingroup%
\def\thefootnote{\fnsymbol{footnote}}\footnote[#1]{#2}\endgroup}
\pgfplotstableset{
    columns={i,Result,i2,Result2},
   columns/i2/.style={
   column name={$i$},
	  },
   columns/i/.style={
   column name={$i$},
	  }, 
   columns/Result2/.style={
   column name={Result},
	  },  
    columns/error1/.style={
        column name=$L_2$,
        sci,sci zerofill,sci subscript,
        precision=3},
    columns/dof/.style={
        int detect,
        column name=\textsc{Dof}},
    every head row/.style={
        before row=\toprule,after row=\midrule},
    every last row/.style={
        after row=\bottomrule}}
\begin{document}
\title{On the Symmetry of Polar Codes for Symmetric Binary-Input Discrete Memoryless Channels}



\setlength{\arraycolsep}{0.0em}

\author
{
\IEEEauthorblockN{Qiming Wang}
\IEEEauthorblockA{
School of Information Science and Technology\\
University of Science and Technology of China\\
Email: hugh1234@mail.ustc.edu.cn}
\and
\IEEEauthorblockN{Liping~Li }
\IEEEauthorblockA{
School of Electronics and Information Engineering\\
Anhui University, China,\\
Email: liping\_li@ahu.edu.cn}
}

\maketitle
\begin{abstract}
In this paper, we study the symmetry of polar codes on symmetric binary-input discrete memoryless channels (B-DMC).
The symmetry property of polar codes is originally pointed out in Arikan's work for general B-DMC channels.
With the symmetry, the output vector $y_1^N$ ($N$ be the block length) can be divided into equivalence classes in
terms of their transition probabilities.
In this paper, we present a new 
frame of analysis on the symmetry of polar codes for B-DMC channels.
Theorems are provided to characterize the symmetries among the received vectors.
With this new perspective, we can fully utilize the property of
the underlying channels and reduce the number of equivalence classes. 
The analysis is applied to binary symmetric
channels and shows a great reduction of the number of equivalence classes compared with the original symmetry setting.
\end{abstract}

\section{Introduction}\label{sec_ref}

Polar codes are invented by Arikan in \cite{arikan_iti09}.
Polar codes are the first class of codes that can achieve the capacity for symmetric binary-input discrete
memoryless channels (B-DMC) with a low complexity. The encoding and decoding (with successive cancellation, SC)
has a complexity of $\mathcal{O}(N \log N)$ \cite{arikan_iti09}.
The binary input alphabet in Arikan seminal work \cite{arikan_iti09} is
later on extended to non-binary input alphabet \cite{sasoglu_09,mori_itw10,pradhan_allerton11}.
The construction of polar codes are reported in
\cite{mori_isit09, telatar_isit11, trifonov_itc12, vardy_polar}
and different procedures are
proposed assuming the original $2 \times 2$ kernel matrix.
Polar codes based on the kernel matrices of size $l \times l$ are studied in \cite{korada_iti10}.
In this paper, we study the original binary-input coding scheme.

Let $W$ be a binary discrete memoryless channel (B-DMC). Let $N$ be the code block length.
The transition probability of the $i$th split channel is determined by
the received vector (with a length $N$), and the previously decoded $i-1$ bits, resulting
in a number of possible outputs of $|\mathcal{Y}|^{N+i}$ where $\mathcal{Y}$ is the output alphet.
In \cite{arikan_iti09}, Arikan showed that if $W$ is symmetric, then the split channel
is also symmetric and the number of possible outputs for the $i$th split channel
can be reduced to $|\mathcal{Y}|^i$. The received vectors with the same transition probability form
an equivalence class. The construction of polar codes in \cite{mori_isit09, telatar_isit11, trifonov_itc12, vardy_polar}
did not fully utilize the property of the symmetry of polar codes as we just describe simply because the number
$|\mathcal{Y}|^i$ is still large for large block lengths.

In this paper, we study the further symmetry of polar codes for B-DMC channels. We show that finding
the number of classes of received vectors with the same transition probability can be transformed
into a problem of finding the number of solutions to a set of equations.
Given the symmetry of the underlying channel $W$, the number of solutions
to the set of linear equations can be greatly smaller than $|\mathcal{Y}|^i$ for the $i$th split channel.
We provide theoretical analysis on how to find the equivalent received vectors.
The exact number of equivalence classes for a set of split channels are also provided in closed form expressions.
The symmetry properties of polar codes in this paper are applied to binary
symmetric channels (BSC). The comparison between the number of equivalence classes obtained in this paper
and the original number $2^i$ for the $i$th split channel shows that our frame of analysis can be conveniently
applied when trying to explore the symmetries of polar codes for potential uses.

Following the notations in \cite{arikan_iti09}, in the paper,
we use $v_1^N$ to represent a row vector with elements $(v_1,v_2,...,v_N)$. We also use
$\mathbf{v}$ to represent the same vector for notational convenience. Given a vector $v_1^N$, the
vector $v_i^j$ is a subvector $(v_i, ..., v_j)$ with $1 \le i,j \le N$. If there is a set $\mathcal{A} \in \{1,2,...,N\}$,
then $v_{\mathcal{A}}$ denotes a subvector with elements in $\{v_i, i \in \mathcal{A}\}$.

The rest of the paper is organized as follows. In Section
\ref{sec_background}, the symmetry of polar coding is introduced.
Section \ref{sec_analysis} introduces
theorems used in this paper for characterizing the equivalence classes.
Then theorems and procedures on how to find the exact number of equivalence classes are presented in
Section \ref{sec_counting}.
Concluding remarks are presented in Section
\ref{sec_con}.
\section{Symmetry of Polar Codes}\label{sec_background}
In this section, we briefly restate the symmetry of polar codes introduced in \cite{arikan_iti09} and present
some of our own notations used in sequel of the paper.

Denote the transition probability of the underlying channel $W$ as $W(y|x)$.
Define
\begin{equation}\label{eq_wn}
W_N(y_1^N|u_1^N) = W^N(y_1^N|u_1^NG_N)
\end{equation}
where $W^N(y_1^N|u_1^NG_N)$ is the $N$-time uses of the underlying channel $W$, the block length
$N = 2^n$, and $G_N = BF^{\otimes n} = F^{\otimes n}B$.
The matrix $B$ is the bit reverse permutation matrix, and $F=\left[\begin{smallmatrix} 1&0 \\ 1&1 \end{smallmatrix}\right]$ \cite{arikan_iti09}.
In this paper we use $G_N = F^{\otimes n}$ directly instead of $G_N =BF^{\otimes n}$.
The only difference $G_N = BF^{\otimes n}$ makes, compared with $G_N = F^{\otimes n}$,  is to perform a permutation on the
received vector $y_1^N$, which doesn't affect our analysis.
Note that throughout this paper, the input vector $u_1^N$, the matrix $F=\left[\begin{smallmatrix} 1&0 \\ 1&1 \end{smallmatrix}\right]$, and the generator matrix $G_N$
are all defined on the binary field $\mathbb{F}_2$. The output $y_1^N$ is defined on the set
$\mathcal{Y}$.

The operation in (\ref{eq_wn}) is the channel combining stage.
In the channel split stage, the $i$th split channel has a transition probability of
\begin{equation}\label{eq_wn_polar}
W_N^{(i)}(y_1^N, u_1^{i-1} | u_i) = \sum_{u_{i+1}^N \in \mathcal{X}^{N-i}}\frac{1}{2^{N-1}}W_N(y_1^N|u_1^N)
\end{equation}
To facilitate the analysis, the operator $\cdot$ is defined as:
\begin{eqnarray} \label{eq_cdot0}
  0 \cdot y &=& y, ~y \in \mathcal{Y} \\ \label{eq_cdot1}
  1 \cdot y &=& v, ~y,v \in \mathcal{Y}
\end{eqnarray}
and $v$ satisfies
\begin{eqnarray}
 W(v|0)&=&W(y|1)\\
W(v|1)&=&W(y|0)
\end{eqnarray}
In other words, $v$ is the corresponding symbol that is symmetric to $y$.
To divide the output alphabet into two groups, we also define two sets:
\begin{eqnarray}
\self(\mathcal{Y}) &=& \{y \in \mathcal{Y}| ~1\cdot y = y\}\label{def_self}\\
\symm(\mathcal{Y}) &=& \{y \in \mathcal{Y}| ~1\cdot y \neq y\}\label{def_symm}
\end{eqnarray}
Set $\self(\mathcal{Y})$ includes the output $y$ which has a symmetric symbol of itself while set $\symm(\mathcal{Y})$
has elements with a different symmetric symbol.
Obviously, the cardinality of the set $\symm(\mathcal{Y})$, $|\symm(\mathcal{Y})|$, is an even number.
As shown in \cite{arikan_iti09}, for symmetric B-DMC channels, it is sufficient to discuss
\begin{equation}\label{eq_wn0}
W_N^{(i)}(y_1^N, 0_1^{i-1} | 0) = \sum_{u_{i+1}^N \in \mathcal{X}^{N-i},u_1^i = 0_1^i }\frac{1}{2^{N-1}}W_N(y_1^N|u_1^N)
\end{equation}
as every $W_N^{(i)}(y_1^N, u_1^{i-1} | u_i)$ is equivalent to a $W_N^{(i)}(\tilde{y}_1^N, 0_1^{i-1} | 0)$ where
$\tilde{y}_1^N$ is selected from the following set:
\begin{equation}\label{eq_xi1}
\tilde{\mathcal{X}}_{i+1} = \{y_1^N: ~ a_1^NG_N \cdot y_1^N, ~ \text{and}~ a_1^i = u_1^i\}
\end{equation}
In (\ref{eq_xi1}), the operation $a_1^NG_N \cdot y_1^N$ is the element-wise operation of the $\cdot$ operation
defined in (\ref{eq_cdot0}) and (\ref{eq_cdot1}).
In this paper, we will simply use $W_N^{(i)}(\mathbf{y}), \mathbf{y}\in \mathcal{Y}^N$
to replace $W_N^{(i)}(y_1^N, 0_1^{i-1} | 0)$ as the previous bits and the current bit are all zeros.
And if two received vectors $\mathbf{y}$ and $\mathbf{v}$ have the same transition probability
 \begin{equation}W_N^{(i)}(\mathbf{y}) = W_N^{(i)}(\mathbf{v})\end{equation}
then we call $\mathbf{y}$ and $\mathbf{v}$ probability equivalent on bit channel $i$.
Seen from (\ref{eq_xi1}), if $\mathbf{y}$ and $\mathbf{v}$ are probability equivalent on bit channel $i$,
they are also probability equivalent on the row space of the submatrix $G_{i+1}^N$,
where $G_{i+1}^N$ means the rows of
$G_N$ from the $(i+1)$th row to the $N$th row.
\section{Characterization of the Symmetry}\label{sec_analysis}
For the convenience of the description, we denote the submatrix $G_{i+1}^N$ of the matrix $G_N$ as $A(N,i)$.
Then $A(2N,i)$ is the submatrix  formed
by the rows from $(i+1)$ to $2N$ of $G_{2N} = F^{\otimes n+1}$.
For a given block length $N$, we abbreviate $A(N,i)=G_{i+1}^N$ as $A = G_{i+1}^N$ since the size of $A$ is clear
from the context.
Denote $\row(A)$ as the row space of $A$.
For the received vector $\mathbf{y}\in \mathcal{Y}^N$, define
\begin{equation} \label{eq_ray}
\row(A) \cdot \mathbf{y} = \{\mathbf{u} \cdot \mathbf{y}|~\mathbf{u} \in \row(A) \}
\end{equation}
In the following, we use a simpler notation to refer to the set defined in (\ref{eq_ray}):
 $\row(A,\mathbf{y}) = \row(A) \cdot \mathbf{y}$.
Seen from (\ref{eq_wn0}), the transition probability for the $i$th channel is exactly summed over the
row space of $A$. Using the notation $W_N^{(i)}(\mathbf{y}) = W_N^{(i)}(\mathbf{y}, 0_1^{i-1}|0)$,
we can rewrite (\ref{eq_wn0}) as
\begin{eqnarray} \label{eq_wny}
W_N^{(i)}(\mathbf{y}) &=& \sum_{\mathbf{u} \in \row(A)}\frac{1}{2^{N-1}}W_N(\mathbf{y}|\mathbf{u})\\\label{eq_wnv}
&=& \sum_{\mathbf{u} \in \row(A)}\frac{1}{2^{N-1}}W_N(\mathbf{u}\cdot \mathbf{y}|0_1^{N})\\\
&=&\sum_{\mathbf{v} \in \row(A,y)}\frac{1}{2^{N-1}}W^N(\mathbf{v}|\mathbf{0})\label{eq_sim}
\end{eqnarray}
The equality in (\ref{eq_wnv}) is obtained from the application of the symmetry property (\ref{eq_cdot0}) and (\ref{eq_cdot1})
of the underlying channel $W$. 

In the following, we present theorems to characterize the properties of the received vectors which belong
to the same equivalence class. With our frame of analysis, the original symmetry property presented in
\cite{arikan_iti09} is also provided.
\newtheorem{theorem}{Theorem}
\newtheorem{corollary}{Corollary}
\begin{theorem}\label{theorem_basic1}
Let $P$ be a permutation matrix. Denote $\T_P(\cdot)$ as the permutation transformation.
If there exists an invertible matrix $H$ so that
$AP = HA$, then $\mathbf{v} = \T_P(\mathbf{y})$ and $\mathbf{y}$ are probability equivalent on $A$, where $A$ is $A(N,i)$.

\end{theorem}
\begin{IEEEproof}
Since the matrix $P$ is a permutation matrix, the inverse matrix of it is $P^T$. Thus the transformation $\T_P$ is a bijection.
Let's consider the two spaces $\row(A,\T_P(\mathbf{y}))$ and $\row(AP, \T_P(\mathbf{y}))$. Obviously the row space of $A$ and $HA$ is
the same, so the row space of $A$ and $AP$ must also be the same, since $AP = HA$ and $H$ is invertible.
For each $\mathbf{z} \in \row(A, \mathbf{y})$, there exists a $\T_P(\mathbf{z}) \in \row(AP, \T_P(\mathbf{y}))=\row(A, \T_P(\mathbf{y}))$ and vice versa.
Since \begin{equation}
W^N(\mathbf{z}|0) = W^N(\textstyle{\T_P(\mathbf{z})}|0)
\end{equation}
$\mathbf{y}$ and $\T_P(\mathbf{y})$ must be probability equivalent on $A$ for the $i$th channel.
\end{IEEEproof}

The following corollary can be easily obtained from Theorem \ref{theorem_basic1} if we assume the matrix $A$ is
an invertible matrix.
\begin{corollary}\label{corollary_basic1}
The vectors $\mathbf{y}, \T_P(\mathbf{y}) \in \mathcal{Y}^N$ are always probability equivalent on
an invertible matrix, where $P$ is a permutation matrix.
\end{corollary}

Note that the matrix $A = G_{i+1}^N$ can't be an invertible matrix for any bit channel $i$. But if $A$ can be
decomposed into several small matrices and each of these small matrices is row equivalent to an identify matrix,
then this Corollary \ref{corollary_basic1} can be used to characterize the symmetry of the received vectors
operated on those small matrices.
\begin{theorem}\label{theorem_basic2}
If there exists a vector $\mathbf{u} \in \mathcal{X}^{N-i}$ so that $\mathbf{k} = \mathbf{u}A$, then
$\mathbf{v} = \mathbf{k} \cdot \mathbf{y}$ and $\mathbf{y}$
are probability equivalent on bit channel $i$.
\end{theorem}

Theorem \ref{theorem_basic2} is equivalent to the symmetry property in \cite{arikan_iti09}. Please refer to \cite{arikan_iti09}
for the proof there.
\begin{corollary}
For BSC channels, the number of equivalence classes for the $i$th bit channel can be calculated from
$\mathbf{y}\in \mathbb{F}_2^N$ and $\mathbf{y}_{i+1}^N=0$.
\end{corollary}
\begin{IEEEproof}
For a BSC channel, the operation in (\ref{eq_cdot0}) and (\ref{eq_cdot1}) becomes
$0 \cdot y = y$ and $1 \cdot y = y \oplus 1$. Also note that columns from $i+1$ to $N$ of the matrix $A$
form an independent set as the matrix $G_N$ is lower triangular.
Therefore, for any $y_1^N$ defined in the binary field $\mathbb{F}_2$, we can find a unique vector
$\mathbf{u} \in \mathcal{X}^{N-i}$ such that
the vector $\mathbf{k} = \mathbf{u}A$ and $k_{i+1}^N = y_{i+1}^N$.
This makes the vector
$\mathbf{v} =\mathbf{u}A \cdot \mathbf{y}$
have  $\mathbf{v}_{i+1}^N=0_1^{N-i}$. From Theorem \ref{theorem_basic2}, it's seen that $\mathbf{v}$
and $\mathbf{y}$ are probability equivalent on the $i$th bit channel. Thus the equivalence classes
for BSC channels can be investigated based on the received vector $v_1^N$ with $v_i^{N-i} = 0_1^{N-i}$, which
results in $2^i$ as the largest possible number of equivalence classes.
\end{IEEEproof}

The following Theorem \ref{theorem_basic3} states that the equivalence of two vectors for the $i$th
bit channels is preserved when the block length increases from $N$ to $2N$.
\begin{theorem}\label{theorem_basic3}
If vector $\mathbf{y}$ and $\mathbf{v}$ are probability equivalent on $A(N,i)=A$, then
$\mathbf{u} = (1,0) \otimes \mathbf{y}$ and $\mathbf{t} = (1,0) \otimes \mathbf{v}$ are also
probability equivalent on $A(2N,i)$.
\end{theorem}
\begin{IEEEproof}
Define $B = (1,0)\otimes A$, $C = A(2N,i)$, and $D = A(2N,N)$.
The matrix $D$ is row equivalent to $(1,1)\otimes I_N$ from the definition of the generator matrix $G_N$.
It's clear that the row space of the matrix $B$, $D$ are subsets of the row space of matrix $C$: $\row(B) \subset \row(C)$, $\row(D) \subset \row(C) $.
If $\mathbf{y}$ and $\mathbf{v}$ are 
probability equivalent on $A$,
there exists a bijection $\T_{P_h}$ such that
\begin{equation}
W^N(\mathbf{h}|\mathbf{0})=W^N(\textstyle{\T_{P_h}}(h)|\mathbf{0})
\end{equation}
where $\mathbf{h}\in R(A, \mathbf{y})$, $\textstyle{\T_{P_h}}(h)\in R(A, \mathbf{v})$, and $P_h$ is a permutation matrix 
that may vary with $\mathbf{h}$.
Then $\mathbf{u}$ and $\mathbf{t}$ must also be probability equivalent on B,
since there exists a bijection $f$ such that
\begin{equation}
f:\mathbf{h}\in R(B, \mathbf{u}) \to \mathbf{h}L_\mathbf{h} \in R(B, \mathbf{t}), ~L_\mathbf{h}= (1,1)\otimes P_\mathbf{h}
\end{equation}
 Now, there's a partition of $R(C, \mathbf{t})$,
\begin{equation}
\partition(\mathbf{t}) = \{\row(D)\cdot \mathbf{r}|~\mathbf{r}\in R(B, \mathbf{t})\}
\end{equation}
 The same for $R(C, \mathbf{u})$,
\begin{equation}
\partition(\mathbf{u}) = \{\row(D)\cdot \mathbf{r}|~\mathbf{r}\in R(B, \mathbf{u})\}
\end{equation}
For any $\mathbf{e}\in R(C, \mathbf{u})$, it must belong to some $\row(D)\cdot \mathbf{r} \in \partition(\mathbf{u})$.
According to Theorem \ref{theorem_basic1}, $\mathbf{r}\in R(B, \mathbf{u})$ and
$f(\mathbf{r})\in R(B, \mathbf{\mathbf{t}})$
 are probability equivalent on $\row(D)$, since $D$ and $DL_\mathbf{h}$ are row equivalent. It follows that there exists a bijection
$\mathbf{e'_r(e)}\in \row(D)\cdot f(\mathbf{r}) \subset R(C, \mathbf{t})$ that satisfies
$\mathbf{e'_r(e)} = \mathbf{e}P_\mathbf{e}$.
Therefore, $\mathbf{u}$ and $\mathbf{t}$ are probability equivalent on A(2N,i).
\end{IEEEproof}
\begin{corollary}
For BSC channels (which means $\mathcal{Y}=\mathbb{F}_2$), the number of different transition probabilities for bit channel $i$
doesn't change with the block length $N$.
Though the value of the transition probability
$W_N^{(i)}(\mathbf{y})$ and $W_{2N}^{(i)}((1,0)\otimes \mathbf{y})$ may not be the same for
different block lengths.
\end{corollary}
\section{Evaluation of the Number of 
Equivalence Classes
}\label{sec_counting}
In Section \ref{sec_analysis}, the symmetry property is introduced which can be used to
determine  if any two outputs have the same transition probability for bit channel $i$.
In this section, we analyze the exact number of different transition probabilities for bit channel $i$.
In \cite{arikan_iti09}, this is the problem to find the number of equivalence classes.
\subsection{Transformation of the Output Alphabet}\label{trans_def}
In this section, we transform the study of the symmetry based on the output $\mathbf{y} \in \mathcal{Y}^N$ to another output
alphabet, which is more convenient to manipulate.

Before 
Theorem \ref{counting1}
is introduced,
some new notations are needed.
We use $(\mathcal{Y}^N,A, \cdot)$ to denote the problem of studying the symmetry based on the transition probability in equation (\ref{eq_sim}).
Denote $S_1=|\self(\mathcal{Y})|$ and $S_2=|\symm(\mathcal{Y})|$.
Remember that $\self(\mathcal{Y})$ and $\symm(\mathcal{Y})$ are defined in (\ref{def_self}) and (\ref{def_symm}), respectively.
Consider the received vector $\mathbf{y} \in \mathcal{Y}^N$ and we count the number of occurrences of all possible symbols
within this received vector.
Suppose $q_t$ is the number of occurrences of the $t$th symbol of $\self(\mathcal{Y})$ and
$e_t$ is the number of occurrences of the $t$th symbol of $\symm(\mathcal{Y})$.
Without loss of generality, assume the corresponding symmetric symbol of the $t$th symbol is its $(S_2+1-t)$th symbol.
That is
\begin{equation}
W(s_t|0)=W(s_{S_2+1-t}|1), s_t\in \symm(\mathcal{Y}),1\leq t\leq S_2
\end{equation}
Define a new mapping as
\begin{equation}\mathbf{y'(y)}=(q_1,...,q_{S_1},e_1,...,e_{S_2}) \end{equation}
It's seen that $\mathbf{y'(y)}$ is well defined in the sense that for any $\mathbf{y}$,
the mapping $\mathbf{y'(y)}$ is unique. Define a new set corresponding to $\mathcal{Y}^N$ as
\setlength{\arraycolsep}{0.0em}
\begin{eqnarray}
\mathcal{Y}' &=& \{\mathbf{y'(y)}|~\mathbf{y} \in \mathcal{Y}^N\}\label{def_y}\\
\self(\mathcal{Y}')&=&\nonumber\\
\{\mathbf{y'(y)}|~e_t &=& e_{S_2+1-t},
 1\leq t\leq S_2, \mathbf{y} \in \mathcal{Y}^N
\}\\
\symm(\mathcal{Y'})&=&\mathcal{Y}'-\self(\mathcal{Y}')\end{eqnarray}
If $\mathbf{z}=(q_1,...,q_{S_1},e_1,...,e_{S_2}) \in \mathcal{Y'}$, define $*$ as
\begin{eqnarray}
0*\mathbf{z} &=& \mathbf{z}\\
1*\mathbf{z} &=& (q_1,...,q_{S_1},e_{S_2},...,e_1)
\end{eqnarray}
Obviously
\begin{equation}\self(\mathcal{Y}')=\{\mathbf{z}\in \mathcal{Y'}
|~\mathbf{z}=1*\mathbf{z}
\}\end{equation}
\newtheorem{lemma}{Lemma}
\begin{corollary}\label{comb_lemma}
From the definition of the set $\mathcal{Y}'$, the cardinality of it,
$|\mathcal{Y}'|$, is the number of solutions to the following equations:
\begin{equation}
\begin{cases} \sum_{1\leq t\leq S_2}{e_t}+\sum_{1\leq t\leq S_1}{q_t} = N\\
e_t \geq 0, q_t \geq 0
\end{cases}
\end{equation}
The number of solutions to this equation is
$|\mathcal{Y}'| = \binom{N+S_1+S_2-1}{S_2+S_1-1}$,
which is proved in the Appendix.
The size of the set $\self(\mathcal{Y}')$,
$|\self(\mathcal{Y}')|$, 
is the number of solutions to the following equations:
\begin{equation}
\begin{cases} \sum_{1\leq t\leq S_2}{e_t}+\sum_{1\leq t\leq S_1}{q_t} = N\\
e_t=e_{S_2+1-t}, 1\leq t\leq S_2  \\
e_t \geq 0, q_t \geq 0
\end{cases}\label{combeq2}
\end{equation}
The number of solutions of (\ref{combeq2}), which we will denote as $\nf(S_1,S_2,N)$ is:
when $S_1 \neq 0$
\begin{equation}|\self(\mathcal{Y}')| =
\sum_{0\leq r\leq N/2}\binom{r+S_2/2-1}{S_2/2-1} \binom{N-2r+S_1-1}{S_1-1}\end{equation}
and when $S_1 = 0$
\begin{equation}|\self(\mathcal{Y}')| = \binom{N/2+S_2/2-1}{S_2/2-1}\end{equation}
Note that, when $N=1$, $|\self(\mathcal{Y}')|=0$. 
The proof of the solution is omitted in this paper due to the space limit.
And the size of the set $\symm(\mathcal{Y}')$ is
\begin{equation}|\symm(\mathcal{Y}')| = |\mathcal{Y}'| - |\self(\mathcal{Y}')|\end{equation}
which will be denoted as $\nm(S_1,S_2,N)$.                                                                  
\end{corollary}

Applying Corollary \ref{comb_lemma} to the BSC channel, we can obtain the following corollary.
\begin{corollary}
If $S_1 = 0$, $S_2 = 2$,
then
$|\self(\mathcal{Y}')| =1$, $|\symm(\mathcal{Y}')| =N$.
\end{corollary}
\subsection{Evaluation of the Number of 
Equivalence Classes
}
In this section, we use the transformed output alphabet in
Section \ref{trans_def} to evaluate the exact number of equivalence classes.
We first provide the following theorem.
\begin{theorem}\label{counting1}
For bit channel $i$, we denote the number of the remaining rows of the generator matrix $G_N$ as
$a = N-i$ ($N = 2^n$).
Let $a'$ be any number in the range $N\geq a'\geq a$ and
$a'=2^{k'}$ for some number $k'$.
Denote the number of different transition probabilities of bit channel $i$ as $\nc(N,i,S_1,S_2)$,
where
$S_1=|\self(\mathcal{Y})|$ and
$S_2=|\symm(\mathcal{Y})|$.
Then the number of different transition probabilities for bit channel $i$ is
equivalent to
$\nc(N,i,S_1,S_2) = \nc(a', i-(N-a'),S_3,S_4)$, where
$S_3 = \nf(S_1,S_2,N/a')$, $S_4 = \nm(S_1,S_2,N/a')$.
\end{theorem}
\begin{IEEEproof}
We know
\begin{equation}
F^{\otimes n}=F^{\otimes n-k'}\otimes F^{\otimes k'}
\end{equation}
The last $a'$ rows are $(1,1,...1)\otimes F^{\otimes k'}$.
Define vector $\mathbf{z}(\mathbf{y}) \in \mathcal{Y}'^{a'}$ as
\begin{equation}
\mathbf{z}(\mathbf{y})\{j\}= \mathbf{y'}(\mathbf{y}\{j,j+a',j+2a'...j+N-a'\}),0<j\leq a'
\end{equation}
where $\mathbf{y}$ is the output vector. Here $\mathbf{y}\{j\}$ means the $j$th entry of $\mathbf{y}$. Look at the definition of $\mathcal{Y}'$,
it's seen that $\mathbf{z}$ is surjective.
For any given $\mathbf{y}$ we can get a unique $\mathbf{z}(\mathbf{y})$.
Also, from the definition of $\mathbf{y'}$, $W_N^{(i)}(\mathbf{y})$ 
can be calculated from 
$\mathbf{z}(\mathbf{y})$.
Observe
\begin{equation}\label{linear_eq}
\mathbf{z}(((\underbrace{1,1,...,1}_{N/a'})\otimes\mathbf{v}) \cdot \mathbf{y})=\mathbf{z}(\mathbf{y})*\mathbf{v}
\end{equation}
where $\mathbf{v}\in \row(A(a',i-(N-a')))$.
Then $(\mathcal{Y}^N,A(N,i), \cdot)$ is equivalent to $(\mathcal{Y'}^{a'},A(a',i-(N-a')), *)$.
So we get $\nc(N,i,S_1,S_2) = \nc(a', i-(N-a'),S_3,S_4)$.
\end{IEEEproof}
\begin{theorem}\label{counting_ret}
\begin{equation}\nc(N,0,S_1,S_2)\leq\binom{N+S_2/2+S_1-1}{S_2/2+S_1-1}\end{equation}
Obviously, this is not the number of transition probabilities for any bit channel
$i$. However, it can be used to deduce the number of transition probabilities for some bit channnels.
\end{theorem}
\begin{IEEEproof}
Since $rank(A(N,0))=N$, obviously:
\begin{equation}\row(A(N,0)) = \mathbb{F}_2^N\end{equation}
Define $S \subset \symm(\mathcal{Y})$ as a set that satisfies
$|S|=S_2/2$, and
$\forall y_1$, $y_2\in S$, $1\cdot y_1\neq y_2$.
Then for any $\mathbf{y}\in \mathcal{Y}^N$, we can have a
$\mathbf{v}=\mathbf{u}\cdot \mathbf{y}, \mathbf{u}\in \mathbb{F}_2^N, \mathbf{v}\in \mathcal{Y}^N$,
so that elements of $\mathbf{v}$ will only choose from $S\cup \self(\mathcal{Y})$.
Since any element has $S_2/2+S_1$ choices and the order doesn't matter,
so there's total $\binom{N+S_2/2+S_1-1}{S_2/2+S_1-1}$ cases at most.
\end{IEEEproof}

For some cases, we can have more accurate results than Theorem \ref{counting_ret}.
Define $0\cdot W_i = log{W(y_i|0)}$,
$1\cdot W_i = log{W(1\cdot y_i|0)}$,
and
$D_i = 1\cdot W_i - 0\cdot W_i$, then we have
\begin{theorem}\label{spi}
If $D_i$s are all different for all $i$, then
\begin{equation}\nc(N,0,S_1,S_2)=\binom{N+S_2/2+S_1-1}{S_2/2+S_1-1}\end{equation}
\end{theorem}
\begin{IEEEproof}
\begin{eqnarray}
log(W^N(\mathbf{y|a}))&=& \sum_{1\leq i\leq N}{a_i\cdot W_i},\mathbf{a}\in \mathbb{F}_2^N\\
&=& log(W^N(\mathbf{y}|0_1^N))+ \sum_{1\leq i\leq N}{a_iD_i}
\end{eqnarray}
If $\mathbf{y}$ and $\mathbf{v}$ are probability equivalent, there exists
$\mathbf{u}\in \mathbb{F}_2^N$
so that
$W^N(\mathbf{u}\cdot \mathbf{y}|0) = W^N(\mathbf{v}|0)$. Define $\mathbf{v_2}=\mathbf{u}\cdot \mathbf{y}$.
We can see that $\mathbf{v_2}$ and $\mathbf{v}$ must have the same set of $\{D_i\}$, which means $\mathbf{v_2}$
must be a permutation of $\mathbf{v}$. This shows that the symmetric properties have been discovered (Theorem \ref{theorem_basic1} and Theorem \ref{theorem_basic2}) are enough
to find all equivalent output vectors.
So the less equal sign in Theorem \ref{counting_ret} can be replaced by an equal sign.
\end{IEEEproof}

Also, if bit channel $i$
can be converted to a channel with
$i'=0$, then we can have an exact evaluation.
\begin{corollary}\label{bsc_nume}
If $S_1 = 0$, $S_2 = 2$,
then for $i = N-a$, where $a$ is power of 2,
\begin{equation}
\nc(N,i,S_1,S_2) = \nc(a,0,1,\frac{N}{a})=\binom{a+\frac{N}{2a}}{\frac{N}{2a}}
\end{equation}
\end{corollary}
This is a numeric result on BSC channels. It provides the exact number of different transition probabilities for bit channel $i$.
For $N=2^n$, we choose $a=2^{\left \lfloor{n/2}\right \rfloor}$ to 
calculate the number of equivalence classes.
When $n$ is even
$\nc(N,i,0,2) = $
\begin{equation}
\binom{a+\frac{N}{2a}}{\frac{N}{2a}}= \frac{(\frac{3}{2}\sqrt{N})!}{(\frac{\sqrt{N}}{2})!(\sqrt{N})!}\approx
(\frac{3}{2})^{\sqrt{N}} 3^{\frac{3}{2}\sqrt{N}}
\end{equation}
and when $n$ is odd
$\nc(N,i,0,2) = $
\begin{equation}
\binom{a+\frac{N}{2a}}{\frac{N}{2a}}=
\frac{(\sqrt{2N})!}{((\sqrt{\frac{N}{2}})!)^2}\approx
(
\frac
{(\sqrt{2N})}
{\sqrt{\frac{N}{2}}}
)^{\sqrt{2N}}=
2^{\sqrt{2N}}
\end{equation}
Although these are quite large numbers, they are still much smaller than $2^{i}=2^{N-a}=2^{N-2^{\left \lfloor{n/2}\right \rfloor}}$ obtained using the symmetry in \cite{arikan_iti09}.

\section{Numerical Results} \label{sec_numerical}
To test Corollary \ref{bsc_nume}, we
numerically calculated the number of different transition probabilities
$W_N^{(i)}(\mathbf{y})$ for all $\mathbf{y}\in \mathbb{F}_2^N$.
The numerically calculated results match the numbers calculated with Corollary \ref{bsc_nume}
up to $i=16$.
For larger $i$s,
the time consumed to calculate the number of different transition probabilities is too long to be practical.
From the table, we see for the BSC channels, the number of equivalence classes can be greatly smaller than
$2^i$ for the $i$th channel.

\pgfplotstabletypesetfile{pgfplotstable.example1.dat}

\section{Conclusion}\label{sec_con}

In this paper, we present a frame of analysis to fully explore the symmetries of polar codes, which is used to 
analyze the smallest number of different transition probabilities for each bit channel.
For some of the bit channels, an exact evaluation of the number of 
different transition probabilities (or the number
of equivalence classes) is provided. The analysis for any given bit channel is still 
under investigation. 

\appendix
\section{First Appendix}
\begin{lemma}\label{comso}
The number of solutions to this equation
\begin{equation}
\begin{cases} \sum_{1\leq t\leq S_2}{e_t}+\sum_{1\leq t\leq S_1}{q_t} = N\\
e_t \geq 0, q_t \geq 0
\end{cases}
\end{equation}
 is $\binom{N+S_1+S_2-1}{S_2+S_1-1}$.
\end{lemma}
\begin{IEEEproof}
The equation can be transformed into
\begin{equation}
\begin{cases} \sum_{1\leq t\leq S_1+S_2}{w_t} = N+S_1+S_2\\
w_t \geq 1
\end{cases}
\end{equation}
which is equivalent to
\begin{equation}\label{simp}
\begin{cases}
1 \leq p_t \leq N+S_1+S_2-1 \\
1\leq t\leq S_1+S_2-1\\
p_{t1}\neq p_{t2}\\
p_1 < p_2 <...<p_{S_1+S_2-1}
\end{cases}
\end{equation}
Then $w_t$ can be seen as
\begin{equation}
\begin{cases}
w_1 = p_1\\
w_2 = p_2-p_1\\
...\\
w_{S_1+S_2-1 }= p_{S_1+S_2-1 }-p_{S_1+S_2-2}\\
w_{S_1+S_2 } = N+S_1+S_2-p_{S_1+S_2-1}
\end{cases}
\end{equation}
The number of solutions to equation (\ref{simp}) 
is equivalent to taking $S_1+S_2-1$ balls out of $N+S_1+S_2-1$ balls without considering the order.
Thus the lemma is proven.
\end{IEEEproof}

\bibliography{../../ref_polar}
\bibliographystyle{IEEEtran}

\end{document}